\documentclass[twocolumn,amsmath,amssymb,pra,showpacs]{revtex4}

\usepackage[T1]{fontenc}
\usepackage[latin1]{inputenc}
\usepackage{graphicx}

\begin{document}
\title[Quantum reflection from thin films]{Quantum reflection of ultracold atoms from thin films, graphene, and semiconductor heterostructures}
\author{$^1$T.~E. Judd, $^2$R.~G. Scott, $^3$A.~M. Martin, $^4$B. Kaczmarek and $^4$T.~M. Fromhold}
\affiliation{$^1$CQ Center for Collective Quantum Phenomena and their Applications, Physikalisches Institut, Eberhard-Karls-Universität Tübingen, Auf der Morgenstelle 14, D-72076 Tübingen, Germany \\
$^2$INO-CNR BEC Center, Universit\`a di Trento, Via Sommarive 14, I-38123 Povo, Italy \\
$^3$School of Physics, University of Melbourne, Parkville, VIC 3010 Australia \\
$^4$Midlands Ultracold Atom Research Centre, University of Nottingham, University Park, Nottingham, NG7 2RD, UK}
\date{\today}

\begin{abstract}
We show that thin dielectric films can be used to enhance the performance of passive atomic mirrors by enabling quantum reflection probabilities of over 90\% for atoms incident at velocities $\sim1\:$mm$\:$s$^{-1}$, achieved in recent experiments. This enhancement is brought about by weakening the Casimir-Polder attraction between the atom and the surface, which induces the quantum reflection. We show that suspended graphene membranes also produce higher quantum reflection probabilities than bulk matter. Temporal changes in the electrical resistance of such membranes, produced as atoms stick to the surface, can be used to monitor the reflection process, non-invasively and in real time. The resistance change allows the reflection probability to be determined purely from electrical measurements without needing to image the reflected atom cloud optically. Finally, we show how perfect atom mirrors may be manufactured from semiconductor heterostructures, which employ an embedded two-dimensional electron gas to tailor the atom-surface interaction and so enhance the reflection by classical means.
\end{abstract}


\maketitle

\section{Introduction}
\label{sec:intro}

Hybrid quantum systems, which combine coherent cold atoms with solid structures, hold great promise for both fundamental science and technological applications. There are, for example, exciting possibilities for creating a new generation of devices to measure precisely, and image spatially, gravitational, electric and magnetic fields. Such devices have a key advantage over laser-based technology because atoms couple much more strongly than photons to gravitational and electromagnetic fields. Mirrors are integral to laser devices and the development of analogous matter-wave reflectors is of similar importance for developing atom optics. Despite considerable progress in manipulating atoms with solids, atom-mirror technology has not become either standardized or commercially available. Classical reflection of atoms has been achieved using light sheets, wire gratings and evanescent waves \cite{bongs,roachmirror,slamaglass,judddiff} but the technology is difficult to implement. Using a passive surface, and relying on quantum reflection, offers greater simplicity and flexibility in terms of potential landscape design. Quantum reflection of alkali atoms from the sharp Casimir-Polder (CP) potential step, found a few micrometres from a solid surface, has been demonstrated but, for Bose-Einstein condensates (BECs), the highest reflection probability seen so far is $\sim 0.7$ \cite{pasquinifull,pasquini2full}. For single atoms, reflection probabilities approaching unity have been observed but only for very light atoms, which approach the surface at extremely acute angles of incidence \cite{shimizu}.

\begin{figure}
\includegraphics[width=1.0\columnwidth]{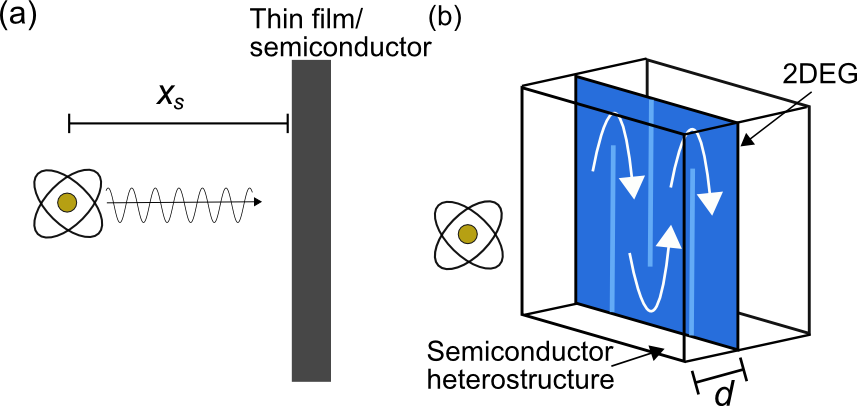}
\caption{\label{fig:schemeref} (a) Schematic diagram (not to scale) showing an alkali atom approaching a dielectric surface (filled rectangle), where $x_s$ is the atom-surface separation. Horizontal arrow indicates the atom's motion towards the surface and the sinusoidal curve represents the atomic matter wave. (b) Schematic diagram showing the relative position and structure of the two-dimensional electron gas (dark blue plane), formed at distance $d$ below the surface of a semiconductor heterostructure. Vertical light blue lines mark the position of ion-implanted insulating channels and the curved arrows indicate the current flow, $I$.}
\end{figure}
Quantum reflection is a phenomenon whereby particles can reflect from a scattering potential in the absence of a classical turning point. It occurs at a sharp potential drop. The condition for strong quantum reflection is $\Phi(k)=(1/k^2)dk/dx_s \sim d\lambda/dx_s \gtrsim 1$ \cite{shimizu} where the local wave number, $k=2\pi/\lambda$, depends on the de Broglie wavelength, $\lambda$, for the atom's centre-of-mass motion, and $x_s$ is the normal distance from the atom to the surface. The mutual van der Waals attraction between an atom and a surface creates a sharp CP potential, which can satisfy the quantum reflection condition. In addition to prospective applications in passive atom-optical mirrors, quantum reflection can also yield information, which is hard to obtain by other means, about the CP potential profile for atoms near both planar and patterned surfaces \cite{pasquinifull,pasquini2full} and about atom-atom interactions within a BEC: for example soliton and vortex production during reflection \cite{scottfull}. 

Here, we show that quantum reflection probabilities can be increased to over 90\% for experimentally-accessible velocities \cite{pasquinifull,pasquini2full} by reducing the thickness of the surface structure. Figure \ref{fig:schemeref}(a) shows a schematic diagram of an alkali atom approaching a thin dielectric surface. We find that a $5\:$nm-thick dielectric film gives the highest levels of quantum reflection. A suspended graphene membrane described using a Dirac electron model also produces quantum reflection probabilities significantly higher than for bulk surfaces, and comparable with the thinnest available dielectric sheets. However, if the graphene membrane is treated within a two-dimensional plasma sheet (hydrodynamic) model \cite{churkin}, the reflection probabilities are lower than for a bulk dielectric. At present there are uncertainties in the application and validity regimes of these models. Our analysis confirms predictions by Churkin \textit{et al.} that present quantum reflection experiments are accurate enough to distinguish between them \cite{churkin}. 

We show that free-standing graphene membranes offer two key advantages over bulk surfaces and dielectric membranes for quantum reflection studies. Firstly, the membrane can be cleaned between successive quantum reflection runs by passing a current through the membrane in order to heat it and so remove adsorbates. Such adsorbates are undesirable because they can strongly affect the atom-surface potential and so complicate the analysis of experimental data \cite{pasquinifull,pasquini2full}. Secondly, the spatio-temporal dynamics of the reflecting cloud, and its reflection probability, can both be determined non-invasively and in real time by measuring changes in the membrane's electrical resistance produced by the adsorption of those atoms that do not quantum reflect. In contrast to previous optical imaging, the electrical measurements that we propose can monitor the time evolution of the atom cloud throughout a single quantum reflection event and are unaffected by fragmentation of the atom cloud that occurs at low approach velocities \cite{pasquinifull,pasquini2full,scottfull}.   

Finally, we explore the possibility of creating perfectly reflecting atom mirrors in which the reflection is enhanced by an array of wires only $\sim 15\:$nm thick, buried in a semiconductor heterostructure. We find that 100\% reflection can be achieved with just $\mu$A wire currents: low enough for the wire array to be formed in a high-quality two-dimensional electron gas (2DEG) \cite{juddZP,sinuco2deg} by using ion-beam implantation to create insulating channels, which define the wires. The architecture of this system is shown in figure \ref{fig:schemeref}(b) where the dark (light) blue areas represent the 2DEG (insulating channels) and the white arrows indicate the current flow.

The paper is structured as follows. In Section \ref{sec:diel}, we investigate pure quantum reflection from thin dielectric glass films. In Section \ref{sec:graph}, we consider quantum reflection from a suspended graphene membrane and quantify the effect that adsorbed atoms have on the membrane's electrical resistance. In Section \ref{sec:2deg}, we explore the effect of a 2DEG wire array on classical and quantum reflection from a gallium arsenide (GaAs) semiconductor surface. Finally, in Section \ref{sec:conc} we draw conclusions and suggest promising directions for further theoretical work and experiments.

\section{Quantum Reflection from Dielectric Films}
\label{sec:diel}

\begin{figure}
\includegraphics[width=1.0\columnwidth]{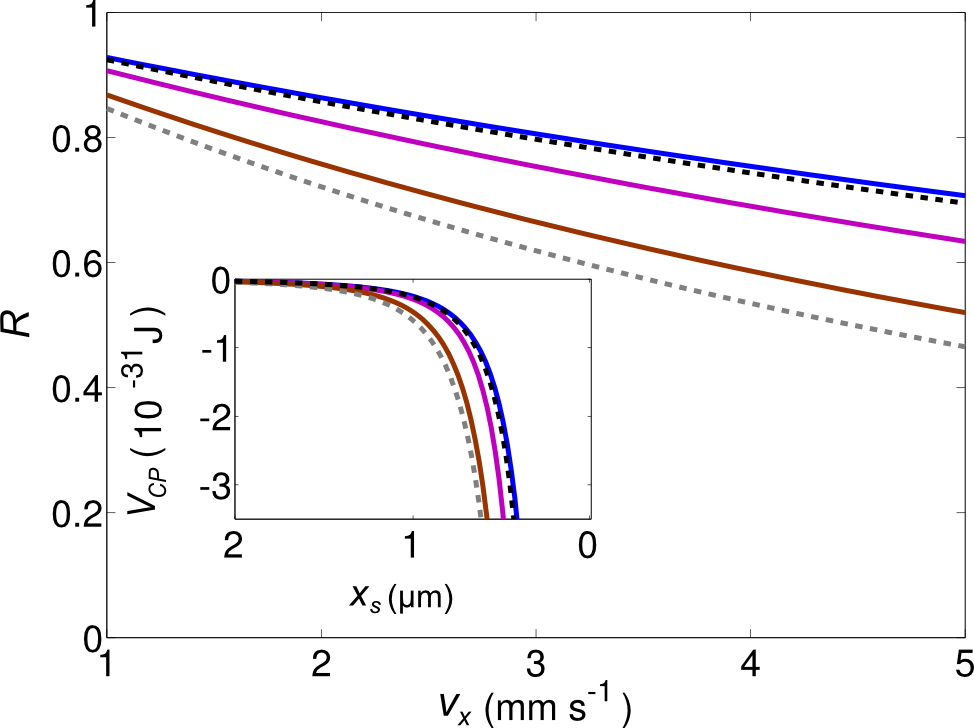}
\caption{\label{fig:Rgraphpthin} Reflection probabilities, $R$, versus incident velocity, $v_x$, for $^7$Li plane waves incident upon (top to bottom) a $5\:$nm dielectric film (blue curve), a graphene monolayer described by the Dirac model (black dashed curve), a $35\:$nm dielectric film (magenta curve), a $10\:\mu$m dielectric film (brown curve), and a graphene monolayer described by the plasma model (grey dashed curve). Inset: Casimir-Polder potentials, colour-coded in the same way as in the main figure.}
\end{figure}

We begin by studying how the properties of thin dielectric films affect the quantum reflection probability, $R$, of a $^7$Li atom approaching the film at normal incidence with velocity $v_x$. The $v_x$ range 1-5$\:$mm$\:$s$^{-1}$ shown in figure \ref{fig:Rgraphpthin} corresponds to that already attained experimentally for ultracold atoms \cite{pasquinifull,pasquini2full}. Lithium-7 is particularly suitable for quantum reflection because it is comparatively light, which ensures maximal $\lambda$ for a given $v_x$, and Bose-condensable \cite{hulet2,hulet}. Further advantages are the low electrical polarizability compared with other alkali atoms, and the tuneability of inter-atomic interactions within a $^7$Li BEC from repulsive to attractive. These factors reduce topological cloud disruption that can occur when a BEC reflects \cite{scottfull,cornishjudd,pasquini2full}, surface overlap, potential screening, and atom losses due to the quantum pressure of the wavepacket \cite{scottfull,cornishjudd}. If these issues can be circumvented, reflection probabilities for BECs are approximately the same as those for plane waves.

We first consider how the thickness, $d_x$, of the dielectric film affects the CP potential, $V_{CP}$, and hence, $R$. We calculate $V_{CP}$ between an atom and the film using the Lifshitz approach \cite{babbcp,reyesdielslab,scheelrev}
\begin{align}
\label{eq:lifslab}
&V_{CP}(x_s) = -\frac{k_B T}{8x_s^3} \sum_{l=0}^{\infty}{}' \alpha(i\xi_l) \int_{2x_s\xi_l/c}^{\infty} ds \: e^{-s} \times \nonumber \\
&\left\{ 2s^2 r_{M}(i\xi_l,s) + \frac{4x_s^2\xi_l^2}{c^2} \left[ r_{E}(i\xi_l,s) - r_{M}(i\xi_l,s) \right] \right\},
\end{align}
where $c$ is the speed of light in a vacuum, $T=300\:$K is the temperature of both the surface and the environment, $\alpha(i\xi_l)=\alpha_0\omega_{671}^2/(\omega_{671}^2 + \xi_l^2)$ approximates the polarizability of $^7$Li at the imaginary Matsubara frequency $\xi_l=2\pi k_B T l/\hbar$ with mode number $l$, $\alpha_0=2.7 \times 10^{-39}\:$Fm$^2$ is the static polarizability, $\omega_{671}=2.8 \times 10^{15}\:$rad$\:$s$^{-1}$ is the transition frequency at a wavelength of $671\:$nm (the dominant Lithium spectral line), and $r_{M}(i\xi_l,s)$ and $r_{E}(i\xi_l,s)$ are, respectively, the transverse magnetic and electric reflection probabilities of the surface. These photonic reflection coefficients are determined at each Matsubara frequency and wave number, $q_l$, which relates to the variable of integration in (\ref{eq:lifslab}) by $s=2x_sq_l$.  The dash ($'$) on the Matsubara sum implies half weight for the $l=0$ term. Other symbols have their usual meaning. For $x_s \ll \hbar c / k_B T \approx 7.6\:\mu$m, the values of $V_{CP}$ obtained at $T=300\:$K are very similar to those at $T=0$. As quantum reflection zones (i.e.\ regions of $x_s$ in which $\Phi(k) > 1$) can extend beyond $1\:\mu$m from the surface, we use the full finite-temperature theory to incorporate small corrections due to black-body radiation interacting with the atoms.

Lifshitz theory was found to be consistent, at the relevant distances, with the results of Bender et al.\ who measured the Casimir-Polder potential between an alkali atom and a glass surface \cite{slamaglass}. Stray electric fields and adatoms were ignored and the good agreement between experiment and calculations of the bare Casimir-Polder potential support this.  Limiting forms of Lifshitz theory were also used to calculate quantum reflection probabilities from a semiconductor surface \cite{pasquinifull,pasquini2full}. For most velocities, good agreement was found with the experimental data. At low velocities ($\lesssim 1.0\:$mm$\:$s$^{-1}$), deviations were attributed to technical issues. When comparing Lifshitz theory with experiment it is sometimes necessary to consider corrections due to effects such as surface roughness \cite{klimrev}. The experiments suggest that such effects can be neglected to a good approximation in our case.

The different types of thin film that we consider have different $r_M(i\xi_l,q_l)$ and $r_E(i\xi_l,q_l)$. In the case of a dielectric slab of width $d_x$, the reflection coefficients are given by \cite{bordag}
\begin{eqnarray}
\label{eq:dielrefprobs}
r_{M}(i\xi_l,q_l) &= \frac{\epsilon^2(i\xi_l)q_l^2-k_l^2}{\epsilon^2(i\xi_l)q_l^2+k_l^2+2q_lk_l\mathrm{coth}(k_l d_x)}, \nonumber \\
r_{E}(i\xi_l,q_l) &= \frac{q_l^2-k_l^2}{q_l^2+k_l^2+2q_lk_l\mathrm{coth}(k_l d_x)}, 
\end{eqnarray}
where
\begin{equation}
\label{eq:kl}
k_l = \sqrt{q_l^2+\frac{\xi_l^2}{c^2}(\epsilon(i\xi_l)-1)}
\end{equation}
is the electromagnetic wave number inside the slab. To evaluate $r_M(i\xi_l,q_l)$ and $r_E(i\xi_l,q_l)$, we also require a dielectric response function $\epsilon(i\xi_l)$. We take bulk SiO$_2$ as this is a good insulator and films can be manufactured down to thicknesses of a few nanometres. The dielectric function of bulk SiO$_2$ may be obtained to good approximation by using the following standard procedure: we use experimental $\epsilon$ values taken at real frequencies \cite{palik} and, for ease of computation, project these data onto the imaginary frequency axis using Kramers-Kronig relations \cite{jackson}.

We begin by considering atomic quantum reflection from dielectric slabs with three thicknesses: $10\:\mu$m, $35\:$nm and $5\:$nm. We calculate $V_{CP}$ using equations (\ref{eq:lifslab})-(\ref{eq:kl}).  
We then use this potential to calculate $R(v_x)$ for incident $^7$Li plane waves using the transfer matrix method \cite{gilmore}. 

In figure \ref{fig:Rgraphpthin}, the solid brown curve (2nd from bottom) shows $R(v_x)$ calculated for a $10\:\mu$m film, which is, in effect, a bulk system since $d_x$ is much greater than the separation (typically $\sim 1\:\mu$m) of the quantum reflection zone (where $\Phi(k)>1$) from the surface. The solid magenta curve (3rd from top) is calculated for a film with $d_x=35\:$nm and the solid blue curve is for a $5\:$nm thick dielectric, which is the thinnest commercially-available SiO$_2$ film we are aware of. In the latter case, $R>0.9$ for $v_x\lesssim 1\:$mm$\:$s$^{-1}$. Comparison of the solid curves in figure \ref{fig:Rgraphpthin} shows that reducing $d_x$ weakens the magnitude of $V_{CP}(x_s)$ (inset) and, hence, increases the reflection probability. The reduction in $V_{CP}(x_s)$ follows from equation \ref{eq:dielrefprobs}; reducing $d_x$ reduces the film's ability to respond to electromagnetic fields \cite{reyesdielslab,scheelrev,parsegianvdw}. When the potential is weaker, the atom sees a sharper potential step. This means that the atomic de Broglie wavelength changes more rapidly in space, and, hence, better fulfils the quantum reflection condition $\Phi(k)>1$. It may also be possible to increase $R$ by tailoring the properties of the atom cloud itself, for example using a $^7$Li BEC with attractive interactions in an optical trap, which enables very low approach speeds to be attained \cite{cornishjudd,jurisch}, or by exploiting thermal equilibrium effects \cite{druzmadroqr}.

Yu et al.\ considered quantum reflection from a thin helium film on a substrate \cite{KleppnerHeQR}. They investigated the role of film thickness and found the opposite trend from the one we do - reflection probabilities were higher with thicker films. This is expected in their case \cite{parsegianvdw} and reveals an important distinction between free-standing films, and those on a substrate.

\section{Quantum Reflection from Suspended Graphene Membranes}
\label{sec:graph}
\subsection{Atom-surface potential and reflection probabilities}

Suspended graphene membranes with diameters of 55 $\mu$m have recently been made \cite{Large}. This is comparable with the diameter of Na-atom BECs with repulsive interactions, used in previous quantum reflection experiments \cite{pasquinifull,pasquini2full}, and much larger than that of BECs with attractive interactions, which have great promise for future quantum reflection studies \cite{cornishjudd}. Graphene is a monolayer film, which suggests that $V_{CP}(x_s)$ will be weak enough to induce strong quantum reflection. Since graphene is a low-dimensional conducting material rather than a dielectric, both the coefficients and the functional form of $V_{CP}(x_s)$ differ from the dielectric case \cite{barashwire,dobsonvdw}. However, as before, $V_{CP}(x_s)$ can be calculated using Lifshitz theory \cite{babbcp}.

Due to graphene's unique electronic structure, a full calculation of its spectroscopic response and, hence, its electromagnetic reflection coefficients, is not a trivial matter. However, a comparison with the above results for dielectric films may be achieved by modelling graphene as a two-dimensional gas of massless Dirac fermions. Such a model has been used previously to calculate dispersion forces due to graphene \cite{churkin}. In this case, $r_E(i\xi_l,q_l)$ and $r_M(i\xi_l,q_l)$ take the form
\begin{eqnarray}
\label{eq:grefprobsdirac}
r_{M}(i\xi_l,q_l) &= \frac{\alpha_{\mathrm{{FS}}} q_l P(\tilde{q})}{2 \tilde{q}^2 + \alpha_{\mathrm{{FS}}} q_l P(\tilde{q})}, \nonumber \\
r_{E}(i\xi_l,q_l) &= \frac{\alpha_{\mathrm{{FS}}} P(\tilde{q})}{2 q_l + \alpha_{\mathrm{{FS}}} P(\tilde{q})}, 
\end{eqnarray}
where 
\begin{equation}
\label{eq:Phi}
P(\tilde{q})=4\left( \tilde{\Delta} + \left(\frac{\tilde{q}^2-4\tilde{\Delta}^2}{2\tilde{q}}\right)\mathrm{arctan}\left(\frac{\tilde{q}}{2\tilde{\Delta}}\right) \right)
\end{equation}
is the polarizability of the electron sheet, $\alpha_{\mathrm{{FS}}}=e^2/\hbar c$ is the fine-structure constant, $\tilde{\Delta}=\Delta/\hbar c$,
\begin{equation}
\label{eq:tildeq}
\tilde{q}=\left[ \frac{v_F^2 s^2}{4x_s^2} + \left(1 - \frac{v_F^2}{c^2}\right)\frac{\xi_l^2}{c^2} \right]^{1/2},
\end{equation}
and $v_F\approx 10^6\:$ms$^{-1}$ is the electron Fermi velocity. We take the gap parameter to be $\Delta=0.1\:$eV \cite{churkin}.

We now use equations (\ref{eq:lifslab}) and (\ref{eq:grefprobsdirac})-(\ref{eq:tildeq}) to calculate the Dirac model potential for graphene and use the transfer matrix method to determine $R(v_x)$ (black dashed curve, 2nd from top in figure \ref{fig:Rgraphpthin}) for incident plane $^7$Li matter waves. The $R$ values for graphene are higher than those for the $35\:$nm thick dielectric film (magenta curve, figure \ref{fig:Rgraphpthin}), but not quite as high as $R$ for the $5\:$nm thick film (blue curve, top in figure \ref{fig:Rgraphpthin}). At first sight this seems surprising since one might expect graphene to produce higher reflection probabilities due to its monolayer thickness. However, graphene's high electron mobility strengthens $\left|V_{CP}\right|$ above that of the $5\:$nm dielectric film (figure \ref{fig:Rgraphpthin} inset) and so lowers $R$.  A similar phenomenon has been observed in experiments \cite{pasquini2full}, which revealed that surfaces with lower electrical conductivity cause greater atomic reflection \cite{footaero}.

As an alternative to the Dirac model, graphene can be modelled by a two-dimensional plasma sheet of electrons \cite{bartonplasma,bordagplasma}. This approach has been used previously for modelling dispersion forces due to graphene and carbon nanotubes \cite{churkin,bordag,blagovgraph}. The plasma model is a simpler theory than the massless Dirac fermion treatment and is expected to be less accurate since it does not incorporate the conical electron dispersion behaviour of Dirac fermions. However, we consider it for comparison because there are uncertainties in the application of the Dirac model. For example, the gap parameter, $\Delta$, is not well known.
The expressions for $r_{M}(i\xi_l,q_l)$ and $r_{E}(i\xi_l,q_l)$ given by the plasma model are
\begin{eqnarray}
\label{eq:grefprobs}
r_{M}(i\xi_l,q_l) &= \frac{c^2 q_l \Omega}{c^2 q_l \Omega + \xi_l^2}, \nonumber \\
r_{E}(i\xi_l,q_l) &= \frac{\Omega}{\Omega + q_l}, 
\end{eqnarray}
where $\Omega= 6.75 \times 10^5\:$m$^{-1}$ is the characteristic electron wave number for the graphene sheet.

We employ equations (\ref{eq:lifslab}) and (\ref{eq:grefprobs}) to calculate $V_{CP}(x_s)$ as before, and the transfer matrix method to calculate $R(v_x)$. Figure \ref{fig:Rgraphpthin} reveals that the $R$ values predicted by the plasma model (grey dashed curve, bottom) are even lower than those for the $10\:\mu$m dielectric film (brown curve) and significantly less than those expected for the $5\:$nm dielectric film (blue curve). The accuracy of $R$ values measured in recent quantum reflection experiments \cite{pasquinifull,pasquini2full,shimizu} is sufficient to distinguish between the Dirac and plasma models as also noted elsewhere \cite{churkin}.

Although the Dirac model of graphene predicts $R$ values (black dashed curve 2nd from top in figure \ref{fig:Rgraphpthin}) slightly below those predicted for the $5\:$nm-thick film (blue curve, top in figure \ref{fig:Rgraphpthin}), graphene offers two major advantages for quantum-reflection studies. Firstly, heating a graphene membrane by passing current through it removes most surface adsorbates \cite{Bolotin}. Such current annealing is particularly effective for cleaning suspended graphene membranes, which, afterwards, can have residual impurity concentrations, $n_{res}$, as low as $\sim 2 \times 10^{14}\:$ m$^{-2}$ \cite {Bolotin}. Consequently, a graphene membrane could be cleaned between successive BEC quantum reflection events to ensure that the incident atoms always interact with a virgin surface. In previous quantum reflection experiments \cite{pasquini2full}, it was suspected that polarization of atomic contaminants on the surface significantly affected the potential landscape of subsequent incoming atoms, thus making it hard to distinguish the effect of the intrinsic CP potential on $R$ from that due to surface adsorbates. 

A second major advantage of using graphene is that the quantum reflection dynamics could be studied, non-invasively and in real time, by measuring changes in the graphene's electrical resistance produced by the adsorption of those atoms that do not quantum reflect. We now consider such resistance changes in detail. 

\subsection{Quantifying the reflection dynamics through changes in the graphene's electrical resistance}

There is considerable interest in alkali-atom adsorbates on graphene and, in particular, their effect on electron transport through the membrane \cite{alk1,alk2,alk3,Naongraphene,alk4,Liongraphene}. Since the adsorbed atoms donate electrons to the graphene and bond ionically to it, they act as positively-charged scattering centres for free electrons within the graphene. As the areal density of adsorbed atoms, $n_{ad}$, increases, the free electron scattering rate and electrical resistivity, $\rho$, of the graphene also increase provided the graphene is sufficiently clean, which requires that the areal density of all the surface impurities $n_{imp}=n_{ad}+n_{res} \lesssim n_{crit} = 2 \times 10^{16}\:$m$^{-2}$ \cite{Maryland}. Consequently, the quantum reflection process can be monitored, and quantified, by measuring the changes in the graphene's electrical resistance, $r$, produced by the adsorption of atoms that do not quantum reflect. 

We now quantify how $r$ changes following quantum reflection of a $^7$Li BEC in which inter-atomic interactions are switched off via a Feshbach resonance, as in previous experiments \cite{hulet}. This BEC contains $N = 3\times 10^{5}$ $^7$Li atoms all in the Gaussian ground state of a harmonic trap with longitudinal (radial) frequencies $\omega_x (\omega_r) = 2\pi \times 3$ $(2\pi \times 193)$ rad s$^{-1}$ \cite{hulet}. The atoms move along the $x$-axis. 
After quantum reflection, the areal density profile of atoms adsorbed on the graphene sheet, which lies in the $y$-$z$ plane, is
\begin{equation}
\label{eq:nad2}
n_{ad}(y,z) = \frac{N[1-R(v_x)]}{\pi l_r^2} e^{-(y^2+z^2)/l_{r}^2},
\end{equation} 
where $l_{r}= (\hbar /m \omega_r)^{1/2}$ is the harmonic oscillator length, in which $m$ is the mass of a single atom. The peak density of the adsorbed atoms, $N(1-R)/(\pi l_r^2)$, attains a maximum value $\sim 1.3 \times 10^{16}\:$m$^{-2}$ when $R=0$ and is therefore $< n_{crit}$ for all $R$. We consider a suspended square graphene membrane with $n_{res} = 2 \times 10^{14}\:$m$^{-2}$, as in recent experiments \cite{Bolotin}, and of sidelength $L = 16$~$\mu$m similar to that reported in \cite{Large}.

\begin{figure}
\includegraphics[width=1.\columnwidth]{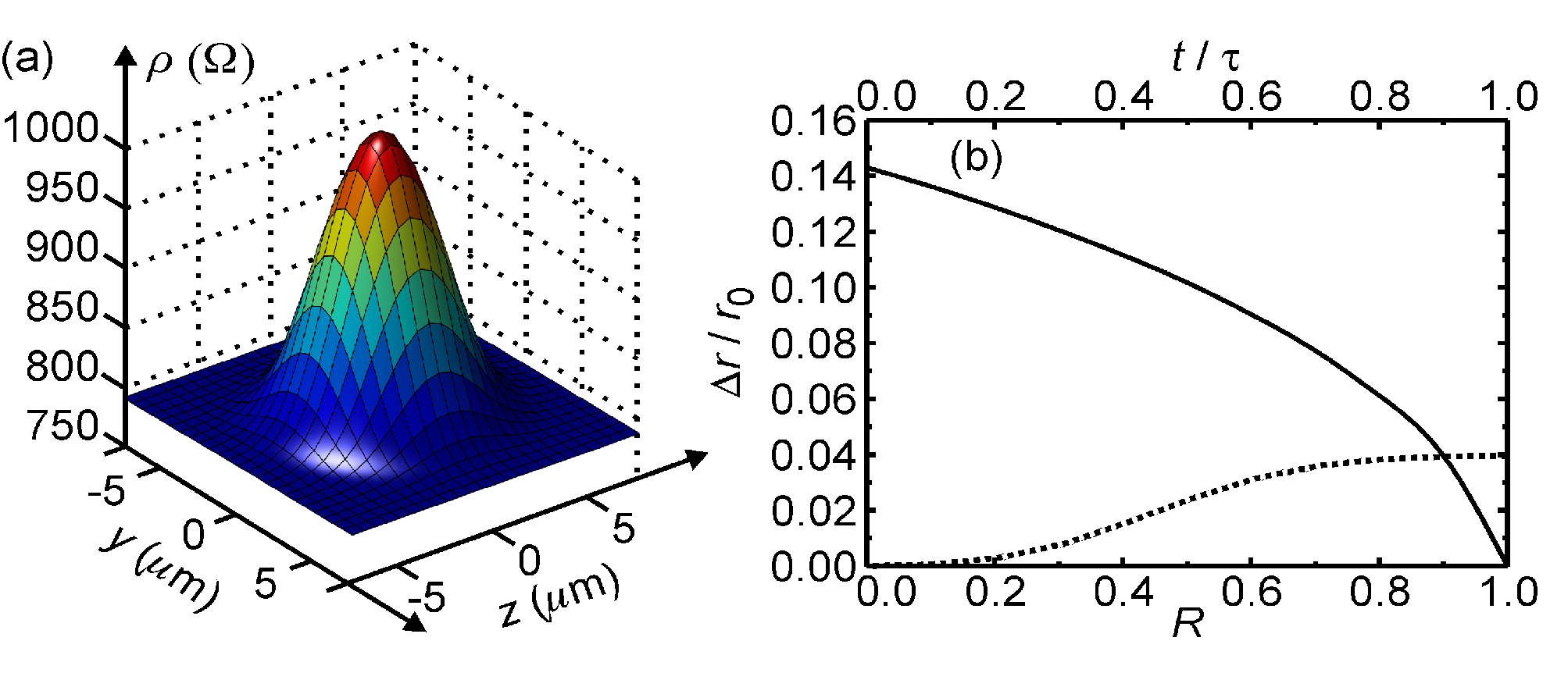}
\caption{\label{fig:fig3} $(a) \rho(y,z)$ calculated for the suspended graphene membrane after quantum reflection of the BEC taking $R=0.9$. (b) Solid curve: $\Delta r /r_0$ versus $R$ (lower horizontal axis) after quantum reflection of the BEC. Dotted curve: $\Delta r /r_0$ versus $t/\tau$ (upper horizontal axis) from the start ($t = 0$) to the end ($t= \tau$) of the BEC's quantum reflection from the graphene membrane, taking $R=0.9$.}
\end{figure}

To determine how the residual and adsorbed impurities affect the local resistivity of the graphene membrane, $\rho(y,z)=1/\sigma(y,z)$, where $\sigma(y,z)$ is the conductivity, we use a model developed in Maryland, specifically equations (1) and (10) of \cite{Maryland}, which agrees well with experimentally-measured variations of $\rho$ with $n_{imp}$ \cite{Maryland,Fuhrer}. Figure \ref{fig:fig3}(a) shows $\rho(y,z)$ calculated after the quantum reflection of the BEC when $R=0.9$, which corresponds to $v_x \sim$ 2 mm s$^{-1}$. Near the edge of the membrane, where $n_{ad} \sim 0$, $\rho(y,z)$ has an approximately constant value $\sim 790\:\Omega$ determined by the value of $n_{res}$. As $n_{ad}$ increases from 0, $\rho(y,z)$ also increases, due to stronger electron scattering, and attains a maximal value $\sim 1000$ $\Omega$ at the centre of the adsorbed Li-atom patch where $(y,z)=(0,0)$. 

To determine how the $\rho(y,z)$ profile shown in figure \ref{fig:fig3}(a) affects $r$, we use a finite-difference method to solve the current continuity equation $\nabla.[\sigma(y,z)\nabla \phi(y,z)]=0$ throughout the membrane subject to the boundary conditions that: (i) the potential $\phi(y=L/2,z)=0$ and $\phi(y=-L/2,z)=V$, where $V$ is a small voltage dropped between ohmic contacts at $y=\pm L/2$; (ii) current flows parallel to the edges of the sample where $z=\pm L/2$. From the potential landscape, $\phi(y,z)$, we calculate the total current 
\begin{equation}
\label{eq:nad}
I = -\int^{L/2}_{-L/2}\sigma(y,z)\left(\frac{\partial \phi(y,z)}{\partial y}\right)dz,
\end{equation}
taking any $y$ value within the sample, and hence determine both $r=V/I$ and the fractional change in the membrane's resistance $\Delta r/r_0$, where $\Delta r = r - r_0$ and $r_0 \sim 790\:\Omega$ is the resistance when $n_{ad}(y,z)=0$. When $R=0.9$, we find $\Delta r/r_0 \sim 0.04$, far above the minimum value of $\sim 10^{-4}$ that can be detected in experiment \cite{detect}. The solid curve in figure \ref{fig:fig3}(b) shows $\Delta r/r_0$ calculated versus $R$ (lower horizontal axis) for the BEC. With increasing $R$, $\Delta r/r_0$ decreases monotonically, meaning that measurements of $\Delta r/r_0$ could be used to determine $R$ independently from the values obtained, in all previous experiments, by optically imaging the BEC before and after quantum reflection. The ability to determine $R$ from electrical resistance measurements alone is a particular advantage at low $v_x$ values where fragmentation of the atom cloud may limit the accuracy of $R$ values obtained by optical imaging \cite{scottfull}. Since $\Delta r/r_0$ values $\gtrsim 10^{-4}$ can be detected in experiment \cite{detect}, the resistance measurements that we propose should be able to measure all $R$ values $\lesssim R_{max} = 0.9995$, and therefore probe quantum reflection over a wide range of $v_x$ extending down to the lowest $v_x$ values considered here and in all previous work. 

The maximum $R$ value that can be measured for the BEC corresponds to the deposition of only $\sim 300$ atoms. Since this detection threshold is so low, real-time measurements of $\Delta r /r_0$ versus $t$ can probe most of the quantum reflection process. To illustrate this, the dotted curve in figure \ref{fig:fig3}(b) shows $\Delta r/r_0$ calculated versus $t/\tau$ (upper horizontal axis) for $R=0.9$, where $\tau = 4l_x/v_x$ is the approximate duration of the reflection process, with $l_{x}= (\hbar /m \omega_x)^{1/2}$. For all $t$ values in this range, $\Delta r /r_0$ exceeds the threshold $\sim 10^{-4}$ required for experimental detection. 

The ability to quantify the number of atoms lost as a function of time during a single quantum reflection event is a key advantage of the electrical measurement scheme that we propose. In previous optical measurements, the time evolution of the reflection process could only be inferred by repeating it many times and imaging the BEC slightly later in each successive run \cite{pasquinifull,pasquini2full}. It may also be possible to probe the reflection dynamics spatially by imprinting an array of nm-scale quantum wires in the graphene membrane, either by etching them or by using local hydrogenation to produce insulating regions between adjacent conducting wires \cite{elias}. Measuring the resistance changes of each quantum wire during quantum reflection would yield spatio-temporal information about the build-up of adsorbed, and hence non-reflecting, atoms on the graphene surface. Resistance-based measurement may be particularly helpful at low velocities since the formation of solitons and vortices in BECs can make accurate determination of $R$ via absorption imaging difficult \cite{scottfull}.  

The change in resistance is most conveniently determined using a small ($\mu V$) AC drive voltage at 200 kHz and measuring the resulting current. The drive frequency is much greater than the inverse reflection time. This prevents the fields due to the current influencing the potential landscape because the field has zero mean on the time scale of the reflection process. In addition, the peak change in potential energy $1\:\mu$m from the surface would be $\sim 10^{-38}\:$J, much less than the atomic energy. Atoms in $m_F=0$ states might be used since they are insensitive to magnetic fields.

Our model should be checked experimentally by comparing the electrical resistance measurements with atom density profiles obtained from optical absorption images. This could be done by moving a cloud of known atom number and density profile slowly against the graphene until all atoms are adsorbed. The measured resistance change could then be checked against our  prediction without needing to image the cloud after interaction with  the surface. Since there are no free parameters in our model, a disagreement would lead us to the important conclusion that our  understanding of the effect of impurities on graphene is incomplete.  Conversely, agreement would validate our claim that resistance measurements can be used to measure reflection probabilities without the use of absorption images.

\section{Quantum and Classical Reflection from Semiconductor Heterostructures}
\label{sec:2deg}

Although we have identified ways to increase quantum reflection probabilities to $> 0.9$, to create ideal atom mirrors the action of the surface must be altered to enable classical reflection of the atoms. Classical reflection from surfaces such as glass prisms has been successfully enhanced by using evanescent waves \cite{slamaglass}. It has also been demonstrated for BECs by using light and magnetic fields \cite{bongs,arnold,andreasdiff2full} to create the potential barrier. In the latter case, a wire array was used as a magnetic mirror. Conventional wires are simple and cheap to make but suffer from Johnson noise \cite{wangtheo,henkelnoisy}, which causes severe atom losses from BECs near surfaces due to atomic spin flips \cite{fortaghsurf,esteve}. Here we show that quantum reflection may be enhanced by using buried wire arrays carrying \textit{extremely low} current. If 100\% reflection can be achieved with currents $I \lesssim 200\:\mu$A, the array can be realised using a two-dimensional electron gas (2DEG) in a semiconductor heterostructure [figure \ref{fig:schemeref}(b)]. Such electron gases have considerably lower Johnson noise than metal wires, owing to fewer defects and higher resistance to thermal currents \cite{crell,wieck,tobben}. Using ion implantation, a high-quality 2DEG wire array can be constructed only $50\:$nm below a solid surface [figure \ref{fig:schemeref}(b)]. This allows cold atoms to approach very close to the conducting elements, thereby strengthening the atom-solid coupling and increasing the performance of such hybrid devices \cite{juddZP}. Metal wire arrays, fabricated on a chip surface, have already been shown to increase reflection of BECs from surfaces \cite{andreas1full}, but their full potential and implementation in 2DEGs has yet to be realised.
\begin{figure}
\includegraphics[width=1.0\columnwidth]{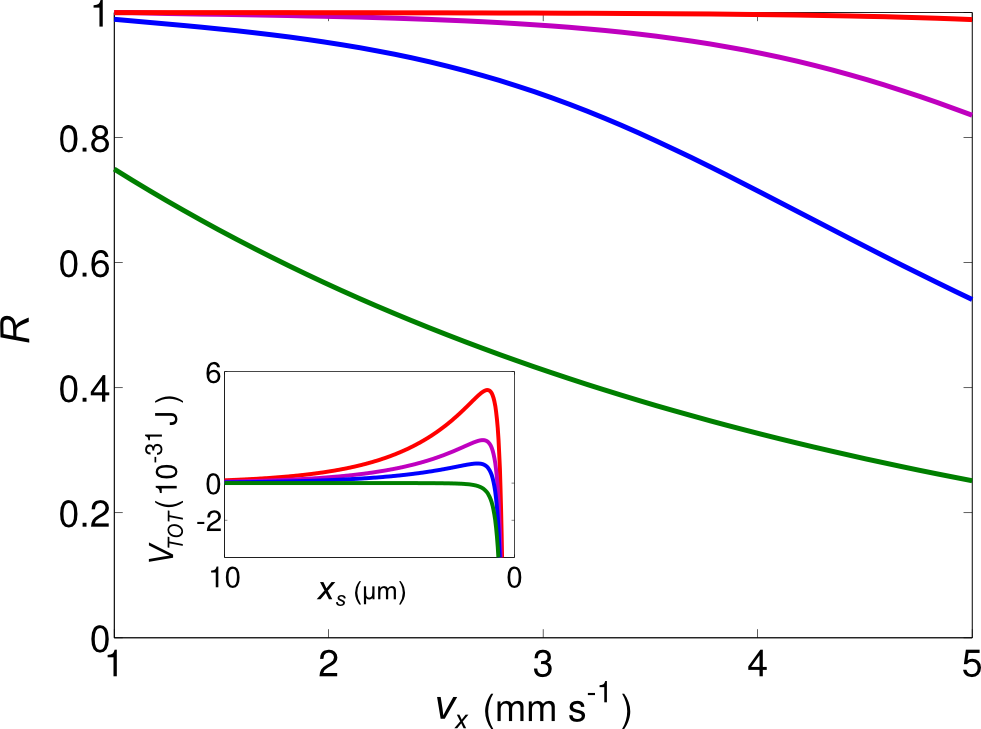}
\caption{\label{fig:marksw2deg} $R(v_x)$ curves for plane Lithium matter waves reflecting from GaAs enhanced by a 2DEG 5$\mu$m below the surface. Currents are (bottom curve to top curve) $0\:\mu$A (green), $1\:\mu$A (blue), $2\:\mu$A (magenta), and $4\:\mu$A (red). Inset shows the total potential $V_{TOT}(x_s)=V_{CP}(x_s)+V_{at}(x_s)$, colour-coded in the same way as the main figure.}
\end{figure}

The magnetic field magnitude at distance $d+x_s$ from the 2DEG plane is given approximately by $B_{app}(x_s) = (2\mu_0 I/b)e^{-2\pi(x_s+d)/b}$ where $b$ is the spatial period of the array and $d=5\:\mu$m is the distance from the 2DEG to the heterostructure surface [figure \ref{fig:schemeref}(b)]. This exponential relation holds provided $x_s + d \gtrsim b/2\pi$. We choose $b = \pi d$ to ensure that the reflecting atoms encounter an almost flat magnetic potential (at $x_s \ll b - d$, the magnetic potential is strongly corrugated). It is important to ensure that $b$ is as large as possible, subject to the requirement that it remains small enough to not significantly corrugate the potential parallel to the surface, since the strength of the potential decreases exponentially as $b$ is reduced. If small corrugations in the potential can be tolerated, larger $b$ values may be chosen. A 2DEG size on the order of $1\:$mm by $1\:$mm, along with the grating period of 5$\pi\:\mu$m, would be large enough to allow the simple exponential description of the potential. See reference \cite{atomchipbookchapter} for further design details.

The potential energy of an atom due to the magnetic field may be written $V_{at}(x_s) = m_F g_L \mu_B B_{app}(x_s)$ where $m_F=2$ defines the electronic spin state and $g_L = 1/2$ is the Land\'e $g$-factor. The exponential decay of $B_{app}(x_s)$ occurs on a length scale $\sim d/2$, comparable with the decay length of the CP potential. Consequently, for sufficiently high $I$ the wire array  creates a sharp potential barrier. In this regime, the magnetic potential can be combined with $V_{CP}(x_s)$ to create a rapidly varying total potential energy, which first increases sharply with increasing $x_s$, attains a positive maximum, and then falls to zero over a few micrometres [figure \ref{fig:marksw2deg}, inset]. This means both quantum and classical reflection can occur.

To demonstrate the effect of the 2DEG current on $R(v_x)$, we calculate $R$ for $^7$Li plane waves reflecting from bulk GaAs  [$V_{CP}(x_s)=-C_4/({x_s} ^{3}(x_s + 3\lambda_a / 2\pi^2))$, $C_4=6.4 \times 10^{-56}\:$J$\:$m$^4$] using the transfer matrix method, but now including $V_{at}(x_s)$ for $I=0$-$4\:\mu$A \cite{foot2degEfield}. The green curve in figure \ref{fig:marksw2deg} (bottom) shows $R(v_x)$ calculated for bulk GaAs with $I=0$ as a reference. Figure \ref{fig:marksw2deg} shows that increasing $I$ to $1\:\mu$A (blue curve, 2nd from bottom) and $2\:\mu$A (magenta curve, 2nd from top) increases $R$ by raising the classical barrier in $V_{TOT}(x_s)=V_{CP}(x_s)+V_{at}(x_s)$ [figure \ref{fig:marksw2deg}, inset]. At $I=4\:\mu$A (red curve, top) the barrier is large enough to produce almost 100\% reflection over the entire range of $v_x$. 

Since the $I$ values that we consider are low enough to be implemented with standard 2DEGs, the proposed scheme offers a practical route to achieving 100\% reflection using well established technology. In addition, because the 2DEG is integrated within the surface, it does not degrade or disrupt the surface, unlike metal surface wires. Small current components across the 2DEG wires, originating from inhomogeneity of the ionised donor distribution, can cause fluctuations in the magnetic field component along the wires. To exponentially suppress such fluctuations, the donor density should be periodically modulated perpendicular to the wires either by optical illumination or by etching narrow (10's of nm) insulating stripes along them \cite{sinuco2deg}.

\section{Conclusions}
\label{sec:conc}

We have shown that atomic quantum reflection probabilities can be increased to $> 0.9$, for typical experimental $v_x$ values, by reducing the thickness of the dielectric substrate and, hence, weakening the CP potential. We have also studied quantum reflection from a suspended graphene monolayer and found that the reflection probabilities exceed those of a bulk dielectric but not those of a $5\:$nm dielectric film, due to graphene's high electron mobility and, hence, stronger CP potential. We have explicitly confirmed predictions that a quantum reflection experiment could be used to distinguish between different models for electron behaviour in graphene \cite{churkin}. We have identified two key advantages of suspended graphene membranes for quantum reflection studies. Firstly, adsorbed alkali atoms can be removed between successive experimental runs by current annealing, thus diminishing the influence of polarized adsorbates on the atoms' potential landscape and reflection probability. Secondly, resistance changes produced by the adsorbed atoms provide a non-invasive real-time route to quantifying the complex spatio-temporal dynamics of quantum reflection. High levels of adsorption might make graphene a better atom reflector than the $5\:$nm glass film since the adsorbates increase the resistance of the sheet and, hence, weaken the potential.

Finally, we have demonstrated that semiconductor heterostructures containing an embedded 2DEG can make perfectly-reflecting atomic mirrors, which should be effective not only for Li but also for heavier atoms, such as commonly-used Rb. Such structures offer exciting possibilities for creating new hybrid quantum systems in which cold atoms couple to solid state devices.

We gratefully acknowledge support from the BW RiSC Programme, the DFG through SFB/TRR21, EPSRC and BW-grid computing resources.


\begin{thebibliography}{56}
\bibitem[{foo({\natexlab{a}})}]{footaero}
\bibinfo{note}{The weaker potential pushes the quantum reflection zone, where
  $\Phi(k) > 1$, closer to the surface. This may explain why quantum reflection
  from aerogels did not produce the theoretically predicted increase of $R$
  \cite{pasquini2full}. Although the potential is very weak, the quantum
  reflection zone spans the same length scale as the surface roughness. This
  probably induces random scattering of the incident atoms (and the photons
  that create $V_{CP}$), thereby reducing $R$.}
\bibitem[{foo({\natexlab{b}})}]{foot2degEfield}
\bibinfo{note}{The electric field due to the voltage applied across the 2DEG
  can be neglected for the small currents we consider here.}
\end{thebibliography}
\end{document}